\documentclass[epj,final]{svjour}
\usepackage{graphics,epsfig,rotating}

\begin{document}

\title{ Peculiarities of isotopic temperatures obtained from
        p + A collisions at 1 GeV }

\author{
     M.N.Andronenko\inst{1}
     L.N.Andronenko\inst{1}
     W.Neubert\inst{2}
and  D.M.Seliverstov\inst{1}
       }
\institute{
    St.Petersburg Nuclear Physics Institute, 188350 Gatchina, Russia \\
    {\em Russian Academy of Science}
\and Institut f\"ur Kern- und Hadronenphysik, FZR inc., 
     01314 Dresden, Germany     
          }

\date{Received: \today}

\abstract{The nuclear temperatures obtained from inclusive measurements
of double isotopic yield
ratios of fragments produced in 1 GeV p+A collisions amount to
$\simeq$ 4 MeV nearly independent from the target mass.  }
 
\maketitle

The pioneering studies in ref.\cite{poch} hint to the transition
between liquid and gaseous phases of nuclear matter. The nuclear temperature
as the crucial observable was derived from the isotope thermometer based on
double yield ratios \cite{albergo} (see below). This thermometer is assumed to
be sensitive to the local temperature at the particle freeze-out \cite{serf}.
Meanwhile, the critical behaviour has been established in various
heavy-ion collisions involving medium and heavy mass nuclei. Whereas the
underlying Statistical Multifragmentation Model \cite{bondorf} was successful in
this mass
region it is open whether the statistical nature of fragmentation processes
can be seen and classified in small many-body systems \cite{botv2}. 
Therefore, temperature measurements in light nuclei with tested 
isotope thermometers are highly desirable.\\
A search for reliable isotope thermometers 
in {\em proton} induced collisions p+Xe$\rightarrow$IMF(3$\le Z \leq 14$)+X
at beam momentum from 80 to 250\,GeV/c 
was recently performed in ref.\cite{tsang} where it was established
that such thermometers show a characteristic behaviour that is {\em
independent} of the reaction type. 
Encouraged by this finding we have analysed the data
available from inclusive measurements of\\
1 GeV proton interactions with various target nuclei.
The data taken into consideration were obtained in some independent
experimental projects performed at the external proton beam of the
PNPI synchrocyclotron Gatchina.\\
(i) One experiment was addressed to Light Charged Particle
(LCP) detection at backward angles \cite{MNA1}: 
\begin{center}
p(1GeV)+($^6$Li,$^7$Li,Be,C,Al,$^{58}$Ni,Ag,Pb)$\rightarrow$LCP(Z=1)+X
\end{center}
The basic tool was a lense spectrometer with a momentum
resolution of $\Delta$p/p$\simeq$2.5\% within the dynamical
range from 0.25 to 0.75 GeV/c. This spectrometer was installed at
$\Theta_{lab}$=109$^o$ and 156$^o$ with respect to the beam axis.
TOF measurements allowed to separate the hydrogen isotopes obtained
from proton collisions with
various targets from $^6$Li to lead. Differential cross sections were
obtained from the kinetic energy spectra extrapolated by fits with a Maxwell 
functional form. 
For the first time, we attend to the yields of hydrogen isotopes
and employed the thermometer based on the double-ratio
($^2$H/$^3$H)/($^1$H/$^2$H).\\
Thus, it became possible to determine
the temperature of $^6$Li as the smallest probe.
Hitherto, the hydrogen isotope yields have casted doubt
on usefulness as thermometer since different nonequilibrium processes may
contribute to these yields. Such contributions should be suppressed under 
our kinematical conditions and we consider the hydrogen yields as an 
adequate tool for temperature measurements.\\
(ii) The second data set to be analysed 
involves Intermediate Mass Fragments (IMF): 
\begin{center}
p(1GeV)+(Be,C,$^{58}$Ni,
Ag,Au,$^{238}$U)$\rightarrow$IMF(Z$\geq$2)+X
\end{center}
As the incident energy was kept fixed at 1 GeV we expect that
target-spectator fragmentation contributes mainly to the observables.
In distinction from heavy-ion collisions the influence of compression
and collective motion on the fragment abundancies 
and on the related temperatures \cite{shlomo}
is supposed to be minimized.\\   
IMF production was studied in p+Ag,\,Au and U collisions
at $\Theta_{lab}$=60$^o$ and 120$^o$ \cite{volnin1}
with a setup consisting of the above mentioned magnetic lense
spectrometer combined with a $\Delta$E-E telescope. The energy resolution
of the $\Delta$E-detector
( $\simeq$ 50 keV) allowed to separate isotopes of fragments
from helium to boron. Absolute cross sections were
obtained by integration of the inclusive energy spectra approximated by a
moving source fit and angular integration using the expression
d$\sigma$/d$\Omega$=$c_1+c_2\cdot cos\Theta_{lab}$.
We included into this analysis additional differential cross sections
at $\Theta_{lab}$=60$^o$ of fragments produced in 1 GeV proton collisions
with $^{48}$Ti, $^{58}$Ni, $^{64}$Ni, $^{112}$Sn and $^{124}$Sn 
\cite{volnin2}.\\     
(iii) Isotopically separated fragments from $^9$Be and $^{12}$C targets
have been registered with an experimental setup consisting of two TOF-E
spectrometers installed at\\ $\Theta_{lab}$=30$^o$ and 126$^o$ with
respect to the beam axis \cite{land1}. The basic detectors in each arm were
twin Bragg Ionization Chambers combined with Parallel Plate Avalanche Counters.
This setup, in general described in ref. \cite{nim}, allowed to measure the 
low energy part of the kinetic energy distributions of the fragments 
below $\simeq$ 30 MeV. This part of the fragment
spectrum, well reproduced by a moving-source fit with {\em one} exponential
slope, is expected to represent mainly the equilibrated component. \\
From the inclusive differential and absolute cross sections 
measured in the mentioned experiments we derived isotopic yield ratios.\\
The method of temperature evaluation from 
isotopic abundancies,ref. \cite{albergo}
is related to five assumptions summarized in ref. \cite{milazzo}. The most
important one is the selection of fragments emitted from a single and
equilibrated source. These conditions are assumed to be fulfilled rather
well at 1 GeV incident energy if the emission from the target spectator is
considered. Experimentally, detection in backward
direction and (or) registration of fragments with low kinetic energies 
should satisfy these requirements. According to ref.\cite{albergo}
the temperature can be obtained from the relation
\begin{equation}
T_{app} =\ \frac B{ln(a \cdot R)}\ ,
\end{equation}
where the double ratio {\em R =R$_1$/R$_2$} is defined by the isotope
yields (Y) 
\begin{center}
R$_1$ = Y(A$_i$,Z$_i$)\,/\,Y(A$_i$+$\Delta$A,Z$_i$+$\Delta$Z) \\
R$_2$ = Y(A$_j$,Z$_j$)\,/\,Y(A$_j$+$\Delta$A,Z$_j$+$\Delta$Z)
\end{center}
which is valid if the fragments with mass A$_i$,A$_j$ and nuclear
charge Z$_i$,Z$_j$ are produced in their ground states.
Each combination of ($R,\,a,\,B$) in equation (1) terms a "thermometer"
which allows to find the absolute or relative temperature related
to the fragment formation. The numerator
{\em B} in equation (1) is determined by the binding energies {\em BE}
\begin{center}
{\em B = BE(A$_i$,Z$_i$)\,-\,BE(A$_i$+$\Delta$A,Z$_i$+$\Delta$Z)\,
    --\,BE(A$_j$,Z$_j$)\,+\,BE(A$_j$+$\Delta$A,Z$_j$+$\Delta$Z)}.
\end{center}
The magnitude $a$ includes the spin degeneration factor and mass
numbers of the considered isotopes. 
The intrinsic nuclear temperature is proportional to the temperature
measured by means of relation (1) up to 5-7 MeV as shown in ref.\cite{hxi}.
We selected pairs with the same
\mbox{ $\Delta$A = $\Delta$Z=1} where the
influence of the chemical potentials cancels out. The choice
of pairs with \mbox{ $\Delta$A =1,\, $\Delta$Z=0} was made to minimize
the influence of Coulomb barriers onto the yields.\\
The relation (1) must be modified for "sequential decays",\,
i.e. if particle decays from higher lying states of the same and other
isotopes contributes to the yields. In the special case of thermometers
selected by $B \geq$ 10 MeV an empirical correction for sequential
decays was published in ref.\,\cite{tsang}
\begin{equation}
\frac 1{T_{app}} = \frac 1{T_o} + \frac{ln(\kappa)}{B}
\end{equation}
where T$_o$ is the unknown intrinsic equilibrium temperature. The
correction factor $\kappa$ is defined by  $R_{app}$=$\kappa \cdot R_o$
where $R_{app}$ is the measured double isotope yield ratio and $R_o$ the
corresponding one for isotopes produced at equilibrium.
The sensitivity of the thermometers improves with
increasing {\em B} reducing relative errors. In the limit
where {\em B} becomes equal or less the intrinsic temperature 
appreciable contributions from sequential decays may affect the yields
\cite{milazzo}.
\begin{figure}
\begin{center}
\epsfig{file=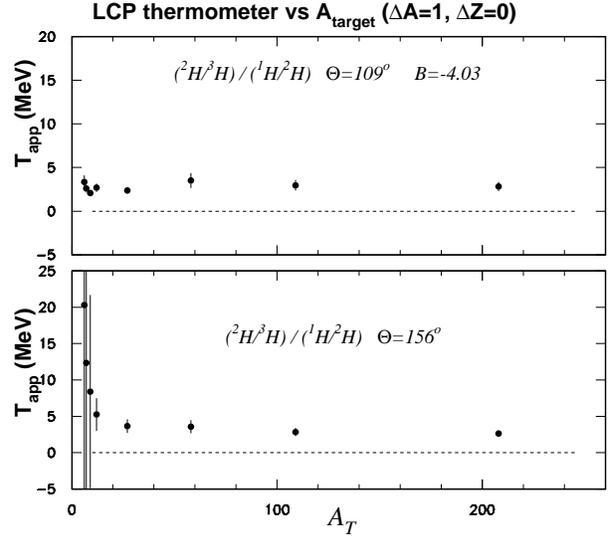,width=8.6cm}
\end{center}
\caption{Apparent temperatures obtained from hydrogen isotopes
as function of the target mass number A$_T$.}
\label{fig:plot1}
\end{figure}
In figs.1--3 we present the dependence of T$_{app}$
on the target mass number A$_T$ as obtained by individual thermometers.
Hereby, neither selection criterion {\em B}$\geq$ 10 MeV 
nor correction for sequential decays were applied to avoid in the first 
instance detailed discussions about it. \\
Fig.\,1 shows the results obtained with the
LCP thermometer ($^2$H/$^3$H)/($^1$H/$^2$H) for two angles.
 One can see that
T$_{app}$ is nearly a constant over the target mass region 6$\leq A_T \leq$ 208
for $\Theta$=109$^o$ (top of fig.1).
Since this behaviour is also established by IMF thermometers (see below) 
we cannot confirm the former doubts about the utility of the
ratio $Y(p)/Y(d)$\,(ref.\cite{albergo}). But the
temperatures which are derived from the hydrogen yields at $\Theta$=
156$^o$ (lower part of fig.1) show some increase toward smaller A$_T$.
We guess that an 
admixture of the $\Delta$ isobaric state becomes apparent in the
used differential cross sections. Fits performed with a
Maxwell-Boltzmann distribution including a Breit-Wigner
contribution evidence this enhancement for the smallest A$_T$ but the
error bars become larger.\\
We plot the temperatures obtained by He and IMF yields as function
of A$_T$ in fig.\,2. Additional data given in fig.\,3  confirm
the finding observed in fig.1 and 2.
Whereas IMF thermometers provide constant temperatures, some
dependence on  A$_T$ is observed if we make use of the ratio $^3$He/$^4$He. 
\begin{figure}
\begin{center}
\epsfig{file=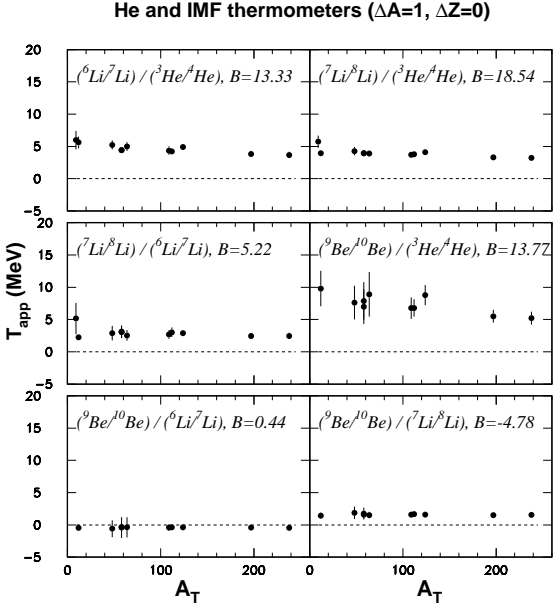,width=8.6cm}
\end{center}
\caption{Temperatures obtained from He and IMF pairs including
 $\Delta$A=1, $\Delta$Z=0 as function of A$_T$.} 
\label{fig:plot2}
\end{figure}
A comparison of temperatures derived from double ratios with 
$\Delta$A=1,$\Delta$Z=0 with those of $\Delta$A=1,$\Delta$Z=1
shows that they are equal within the error bars apart from
some larger fluctuations. The observed agreement of the studied thermometers
exhibits that possibly each one is suitable for relative temperature measurements
without the hitherto introduced limitation {\em B} $\geq$ 10 MeV.
For a given thermometer the influence of
sequential decays seems to be independent from the origin of the
excited primordial fragments. This behaviour is rather surprising since the 
target mass number (or the volume of the fragmenting nuclei, respectively)
changes nearly $\simeq$25 times. Under the same
conditions, the {\em single} ratios of isotope yields show a pronounced
dependence on N$_T$/Z$_T$, ref.\cite{rab}.\\
\begin{figure}
\begin{center}
\epsfig{file=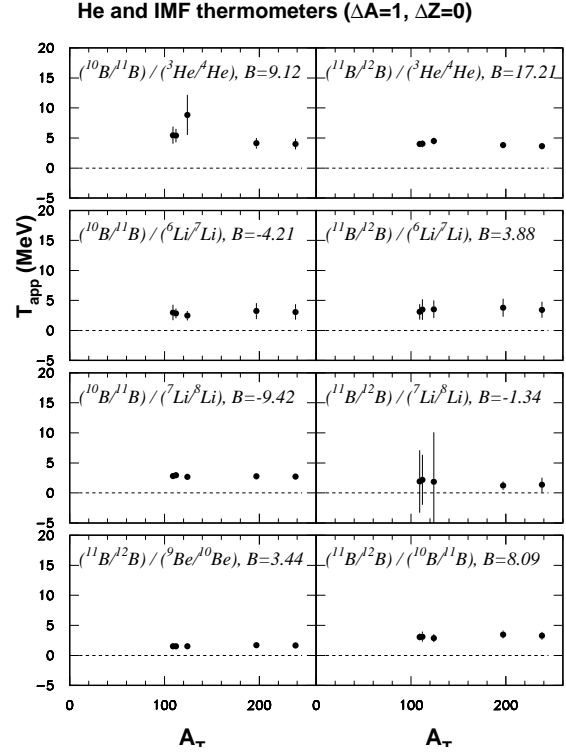,width=8.6cm}
\end{center}
\caption{Temperatures obtained from IMF pairs including combinations
of boron isotopes with those of He,Li and Be.}  
\label{fig:plot3}
\end{figure}
Next we converted the above values T$_{app}$ on the basis of relation (2) 
into intrinsic temperatures T$_o$ as far as the correction factors 
for sequential decay $\kappa$ \cite{hxi} were available.
The mean correction amount to $\simeq$ 5\% but does not exceed 15\%.
Alternative correction methods, e.g. \cite{bon2},
valid in a multifragmentation scenario cannot be applied since contributions
from this process are $\leq$5\% at 1 GeV incident energy.  
The behaviour of individual thermometers is demonstrated in the top panel of
fig.4. Although the drawn errors are overestimated
\footnote{ The errors drawn in figs.1--4 were obtained by simulations
where the primary yields $Y_i \pm \Delta Y_i$ were treated as Gaussian
distributions with $\langle Y_i \rangle$ and $\sigma_i$=$\Delta Y_i$.
Such procedure provides dependable but enlarged errors 
$\Delta T_o$ because it takes not into account that systematic errors
are to be reduced in the ratios $Y_i/Y_{i+1}$ if they are taken from 
the same experiment.}  
the trend observed with one thermometer is repeated by each of the other ones.
Such property hints to a real physical effect. In the lower panel we present
the mean averages of the above values in order to compare with other
available data. The following features of fig.4 are
worth to discuss:\\ 
(i) All temperatures which
have been corrected for sequential decays by using equation (2)
almost coincide at each target mass number A$_T$.
Only the thermometer which explores the double ratio 
($^{11}$B/$^{12}$B)/($^3$He/$^4$He) overestimates the temperature
in the case of the target nuclei $Au$ and $U$.\\
\begin{figure}
\begin{center}
\epsfig{file=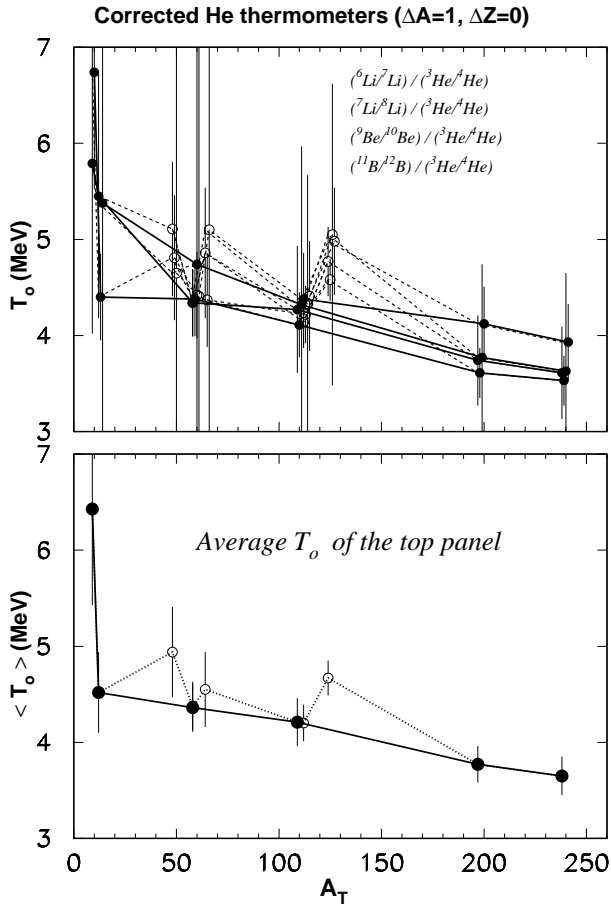,width=8.6cm}
\end{center}
\caption{Corrected temperatures T$_o$ as function of A$_T$.
         Open circles: data from measurements at $\Theta_{lab}$=60$^o$.
         Top panel: individual thermometers are connected by 
         dashed lines, solid lines connect T$_o$ evaluated
         from isotope production cross sections (black dots).  
         Lower panel: mean average of the above 4 thermometers, same
         denotations.}
\label{fig:plot4}
\end{figure}
(ii) The temperatures which have been derived from the differential
cross sections at $\Theta_{lab}$=60$^o$
are larger in comparison with those obtained from production cross sections.
Enhanced temperatures in forward direction and strong variations in the
thermometers with $^3$He/$^4$He ratios were also reported in ref.\,\cite{Vio98}.\\
(iii) Further, we observe pronounced structures in the temperatures.
These irregularities are related to fragments from
the target nuclei $^{48}$Ti,$^{58}$Ni,$^{64}$Ni,$^{112}$Sn and $^{124}$Sn
which were studied to search for the influence of various nucleon composition
on the fragmentation process. 
The authors of this investigation \cite{volnin2} stressed that the yields of 
fragments (except $^4$He) normalized to the geometrical cross section depend
mainly on the $N/Z$ ratio.  
We presume from 
this observation that the fluctuations of the temperature usually attributed
to sequential decays have a causal connection with the nucleon composition
and with the nuclear structure rather than with the size of the fragmenting system.
Such hypothesis is suggested by comparing $\langle T_o \rangle$ of pairs $^{58}$Ni--
$^{64}$Ni and $^{112}$Sn--$^{124}$Sn, respectively.
It seems to be a property of a fragmentation scenario like the
Fermi break-up, ref.\cite{botv1}. \\
(iiii) The temperatures obtained from the ratios of cross sections
show a smooth but weak dependence on the target mass A$_T$. The most pronounced increase
with decreasing A$_T$ was observed only for the thermometers which
include the ratio $^3$He/$^4$He (see fig.4). Above all, we attribute this slope to the
variation of the inherent nucleon constitution of the target nuclei.
Double ratios involving heavier isotopes provide temperatures which
are nearly independent on the target mass (see figs.2--3) within the error bars.
Although hints to this behaviour were already found in 
refs.\cite{tsang},\cite{Vio98} one may doubt the universal
validity of temperature measurements by double isotope ratios.
Therefore, we accomplished an independent test of this method using data from 
an {\em other} physical process. For this purpose,
yields of isotopes from hydrogen to boron registered
in the spontaneous and thermal-neutron induced ternary fission,
\cite{vorob}\cite{krog}, were processed by the same procedure as applied
to them from fragmentation at \mbox{1 GeV.} \footnote
{A forthcoming paper is in progress.}
The used thermometers show a consistent behaviour
resulting in significant lower temperatures of \mbox{
$\langle T_{app} \rangle \simeq$ 1 MeV.} Remark that fission neutron spectra
are well reproduced by a temperature of $\simeq$0.7 MeV
\cite{bowman}.\\
Summarizing, we analyzed inclusive data obtained in\\ \mbox{1 GeV} proton interactions
with various target nuclei employing different isotope thermometers.
We found that even uncorrected thermometers which involve pairs with
{\em B}$\leq$ 10 MeV
provide "stable" results which may be suitable for relative temperature measurements.
The weak dependence of the temperatures from the target mass A$_T$  
suggests speculations about an unique thermodynamical behaviour
whereby the dimensions of the nuclear systems may be changed to a large
extent.\\  
{\em Acknowledgments}
~This work was supported by the German Ministry of Education and 
Research (BMBF) under contract RUS-622-96 and by the Russian Foundation
for Fundamental Research Grant No. 95-02-03671.    


\begin{thebibliography}{99}
\bibitem{poch}
J.Pochodzalla et al.,Phys.\,Rev.\,Lett.\,{\bf 75}, 1040 (1995) 

\bibitem{albergo}
S.Albergo et al., Nuovo\,Cimento\,{\bf 89}, 1 (1985)

\bibitem{serf}
V.Serfling et al., ALADIN collaboration, GSI 98-06,
Darmstadt 1998

\bibitem{bondorf}
J.Bondorf et al., Phys.\,Rep.\,{\bf 257}, 133 (1995)

\bibitem{botv2}
A.S.Botvina and D.H.E.Gross, Phys.\,Rev.\,C\,{\bf 58}, R23 (1998)

\bibitem{tsang}
M.B.Tsang et al., Phys.\,Rev.\,Lett.\,{\bf 78}, 3836 (1997)

\bibitem{MNA1}
M.N.Andronenko et al., Preprint LNPI No. 698 (1981),
Preprint LNPI No. 830 (1983), Preprint LNPI No. 951 (1984) and
Pisma ZhETF\,{\bf 37} 446 (1983) 
\bibitem{shlomo}
S.Shlomo et al., Phys.\,Rev.\,C\,{\bf 55}, R 2155 (1997)

\bibitem{volnin1}
E.N.Volnin et al., Preprint LNPI No. 101 (1974) 

\bibitem{volnin2}
E.N.Volnin et al.,, Phys.\,Rev.\,Lett.\,B\,{\bf 55}, 409 (1975)
and E.N.Volnin, PhD thesis, Leningrad 1975. 

\bibitem{land1}
L.N.Andronenko et al., Preprint PNPI No. 2217 (1998)
and Preprint PNPI No. 2321 (1999)

\bibitem{nim}
L.N.Andronenko et al., Nucl.\,Instr.\,Meth.\,A\,{\bf 312} 467 (1992)

\bibitem{milazzo}
M.Milazzo et al., Phys.\,Rev.\,C\,{\bf 58} 953 (1998) 
 
\bibitem{hxi}
H.Xi et al., Phys.\,Rev.\,C\,{\bf 59}, 1567 (1999)

\bibitem{rab}
L.N.Andronenko et al., in {\em Proceedings of the 7\,$^{th}$
International Conference on Clustering Aspects of Nuclear Structure and 
Dynamics, Rab, 1999}~~to be edited by Z.Basrak et al.,
(World Scientific,Singapore)

\bibitem{bon2}
J.P.Bondorf et al., Phys.\,Rev.\,C\,{\bf 58}, R27 (1998)

\bibitem{Vio98} 
V.E.Viola et al.,Indiana State University IUCF-40007-116,
\,1998 (unpublished)

\bibitem{botv1}
A.S.Botvina et al., Nucl.\,Phys.\,A\,{\bf 475} 663 (1987)  
 
\bibitem{vorob}
A.A.Vorobyov et al., Phys.\,Lett.\,B\,{\bf 30} 332 (1969), 
Phys.\,Lett.B {\bf 40} 102 (1972)

\bibitem{krog}
T.Krogulski et al., Nucl.\,Phys.\,A\,{\bf 128} 219 (1969) 

\bibitem{bowman}
H.R.Bowman et al., Phys.\,Rev.\,{\bf 126} 2110 (1962) 

\end{thebibliography}
\end{document}